\documentclass{emulateapj}
\usepackage{natbib}

\newcommand{\rh}{\varrho}
\renewcommand{\vec}[1]{\mbox{\boldmath$#1$}}

\begin{document}

\title{Transport of toroidal magnetic field by the meridional flow at the
base of the solar convection zone}

\author{Matthias Rempel}

\affil{High Altitude Observatory,
       National Center for Atmospheric Research\footnote{The National
       Center for Atmospheric Research is sponsored by the National
       Science Foundation} , 
       P.O. Box 3000, Boulder, Colorado 80307, USA
      }

\email{rempel@hao.ucar.edu}

\shorttitle{Transport of magnetic field by meridional flow}
\shortauthors{M. Rempel}

\begin{abstract}
In this paper we discuss the transport of toroidal magnetic field
by a weak meridional flow at the base of the convection zone. 
We utilize the differential rotation and meridional flow 
model developed by Rempel and incorporate feedback of a purely toroidal 
magnetic field in two ways: directly through the Lorentz force (magnetic 
tension) and indirectly through quenching of the turbulent viscosity, which
affects the parametrized turbulent angular momentum transport in the model.
In the case of direct Lorentz force feedback we find that a meridional flow
with an amplitude of around $2\,\mbox{m}\,\mbox{s}^{-1}$ can transport a 
magnetic field with a strength of $20\,\mbox{kG}$ to $30\,\mbox{kG}$. Quenching
of turbulent viscosity leads to deflection of the meridional flow from the
magnetized region and a significant reduction of the transport velocity
if the magnetic field is above equipartition strength.
\end{abstract}


\section{Introduction}
Flux-transport dynamos have proven to be successful
for modeling the evolution of the large scale solar magnetic field
\citep{Choudhuri:etal:1995,Dikpati:Charbonneau:1999}. 
In a flux-transport dynamo the 
equatorward propagation of the magnetic activity belt (butterfly diagram)
is a consequence of the equatorward transport of magnetic field at the
base of the convection zone by the meridional flow. 

However, all studies so far addressed the transport of magnetic field by the
meridional circulation in a purely kinematic regime. The toroidal field
strength at the base of the solar convection zone inferred from studies
of rising magnetic flux tubes \citep{Choudhuri:Gilman:1987,Fan:etal:1993,
Schuessler:etal:1994,Caligari:etal:1995,Caligari:etal:1998}
is around $100\,\mbox{kG}$ and thus orders of magnitude larger than the 
equipartition field strength estimated from a meridional flow velocity of a 
few $\mbox{m}\,\mbox{s}^{-1}$. Therefore it is crucial for flux-transport
dynamos to address the feedback of the Lorentz force on the meridional flow. 

In order to be able to address this question it is necessary to incorporate
a model for the solar differential rotation and meridional flow into a
dynamo model and allow for the feedback of the Lorentz force on differential
rotation and meridional flow. Differential rotation and meridional flow
have been addressed in the past through mainly two approaches: 3D full
spherical shell simulations \citep{Glatzmaier:Gilman:1982,Gilman:Miller:1986,
Miesch:etal:2000,Brun:Toomre:2002} and axisymmetric mean field models
\citep{Kitchatinov:Ruediger:1993,Kitchatinov:Ruediger:1995,Ruediger:etal:1998,
Kueker:Stix:2001}. While the 3D simulations have trouble reproducing a
consistent large scale meridional flow pattern (poleward in the upper half 
of the convection zone), as it is observed by helioseismology 
\citep{Braun:Fan:1998,Haber:etal:2002,Zhao:Kosovichev:2004}, such a 
flow is a common feature in most of the mean field models.

In this paper we present a first step toward building a dynamical dynamo
model by focusing primarily on the transport of a prescribed toroidal 
magnetic field by meridional circulation. We do not 
attempt in this very first model to solve the dynamo equations
in order to generate the magnetic fields.
To this end we will use the mean field model for 
differential rotation and meridional circulation described by 
\citet{Rempel:2005} and extend it by incorporating the magnetic tension
resulting from a purely toroidal magnetic field at the base of the solar
convection zone. In this paper we assume that the toroidal magnetic field
consists of a homogeneous layer. We will briefly discuss in section 
\ref{model} how an intermittent field structure will change the results. 
In a future paper we will present a full axisymmetric
mean field dynamo simulation.

\section{Model}
\label{model}
In this paper we utilize the differential rotation / meridional circulation
model of \citet{Rempel:2005} and add the Lorentz force of a toroidal magnetic
field. For the details of the model we refer to \citet{Rempel:2005}. The 
equations we solve are:
\begin{eqnarray}
   \frac{\partial \rh_1}{\partial t} &=& -\frac{1}{r^2}
      \frac{\partial}{\partial r}\left(r^2 v_r\rh_0\right)
      -\frac{1}{r\sin\theta}\frac{\partial}{\partial \theta}
      \left(\sin\theta v_{\theta}\rh_0\right)\label{dens}\\
   \frac{\partial v_r}{\partial t} &=& -v_r\frac{\partial v_r}{\partial r}
      -\frac{v_{\theta}}{r}\frac{\partial v_r}{\partial \theta}
      +\frac{v_{\theta}^2}{r}
      -\frac{\partial}{\partial r} \frac{p_{\rm tot}}{\rh_0}+
      \frac{p_{\rm mag}}{\gamma p_0}g+\frac{s_1}{\gamma}g  \nonumber \\
      && +\left(2\Omega_0\Omega_1+\Omega_1^2\right)r\sin^2\theta 
      +\frac{{F}_r}{\rh_0}-\frac{B^2}{\mu_0\rh_0 r}
      \label{vrad}\\
   \frac{\partial v_{\theta}}{\partial t} &=& 
      -v_r\frac{\partial v_{\theta}}{\partial r}
      -\frac{v_{\theta}}{r}\frac{\partial v_{\theta}}{\partial \theta}
      -\frac{v_r v_{\theta}}{r}
      -\frac{1}{r}\frac{\partial}{\partial\theta}\frac{p_{\rm tot}}{\rh_0}
      \nonumber \\
      && +\left(2\Omega_0\Omega_1+\Omega_1^2\right)r\sin\theta\cos\theta 
      +\frac{{F}_{\theta}}{\rh_0} \nonumber \\
      &&-\frac{B^2}{\mu_0\rh_0 r}\cot\theta
      \label{vthe} \\
   \frac{\partial \Omega_1}{\partial t} &=& 
      -\frac{v_r}{r^2}\frac{\partial}{\partial r}\left[r^2(\Omega_0+\Omega_1)
      \right]\nonumber\\
      &&-\frac{v_{\theta}}{r \sin^2\theta}\frac{\partial}{\partial \theta}
      \left[\sin^2\theta(\Omega_0+\Omega_1)\right]
      +\frac{{F}_{\phi}}{\rh_0 r\sin\theta} \label{omeg}\\
   \frac{\partial s_1}{\partial t} &=& -v_r\frac{\partial s_1}{\partial r}
      -\frac{v_{\theta}}{r}\frac{\partial s_1}{\partial \theta}
      +v_r\frac{\gamma\delta}{H_p}+\frac{\gamma-1}{p_0}{Q}
      \nonumber\\
      &&+\frac{1}{\rh_0 T_0}\vec{\nabla}\cdot(\kappa_t\rh_0 T_0
      \vec{\nabla} s_1)
      \label{entr}
\end{eqnarray}
where 
\begin{eqnarray}
  p_{\rm mag}&=&\frac{B^2}{2\mu_0}\\
  p_{\rm tot}&=&p_1+p_{\rm mag}\\
  H_p&=&\frac{p_0}{\rh_0 g}
\end{eqnarray}
$p_1$ denotes the pressure perturbation with respect to the reference state,
$p_0$, and $s_1$ is the entropy perturbation $s_1=p_1/p_0-\gamma\rh_1/\rh_0$ 
(made dimensionless with the heat capacity $c_v$), 
where $\rh_0$ denotes the density of the reference state.
We use a reference state corresponding to a polytropic atmosphere with gravity
varying $\sim r^{-2}$. Since our model describes the convection zone and 
overshoot region where the deviation from adiabaticity are small 
($\vert \nabla-\nabla_{\rm ad}\vert\ll 1$) we use an adiabatic polytrope
for the reference state. However small perturbations from adiabaticity are 
considered in the entropy equation Eq. (\ref{entr}) through the third term
on the right hand side. We use values for 
$\delta=\nabla-\nabla_{\rm ad}\sim -10^{-5}$ 
below $r=0.725\,R_{\odot}$ and $\delta=0$ above. Different
profiles of $\delta$ influence primarily the differential rotation profile
(more specific: the deviation from the Taylor-Proudman state); however, the 
influence on the meridional flow, which is of primary interest here, is 
very weak. $\Omega_0$ denotes the core rotation rate. $\kappa_t$ in Eq. 
(\ref{entr}) denotes the turbulent convective heat diffusivity and ${Q}$ the
viscous heating. We write the pressure/buoyancy term in Eq. (\ref{vrad}) 
assuming small deviations from adiabaticity 
($\vert \nabla-\nabla_{\rm ad}\vert\ll 1$).

${F}_r$, ${F}_{\theta}$, and ${F}_{\phi}$ denote the viscous stress including 
a parameterization of turbulent angular momentum transport ($\Lambda$-effect,
\citet{Kitchatinov:Ruediger:1995}). The turbulent angular momentum transport
is the driver for the differential rotation and the meridional flow in this
model. The exact form of these terms is discussed in detail in 
\citet{Rempel:2005}.

In this model we do not include the induction equation, which would be 
required to address the non-linear evolution of the field transported by the
meridional flow. As a first step towards the full problem we solve here for
a stationary solution similar to \citet{Rempel:2005}, but include the magnetic
tension force of a toroidal magnetic field. This allows determination of the
field strength up to which a equatorward transport of toroidal magnetic field
is possible if the feedback of the magnetic tension force on the meridional
flow is included. For reasons of simplicity, we omit the magnetic buoyancy 
term in Eq. (\ref{vrad}) (fifth term on right-hand side) and focus this 
discussion only on the magnetic tension force. Magnetic buoyancy has been
addressed in great detail by \citet{Moreno-Insertis:etal:1992} and 
\citet{Rempel:etal:2000}.

In this study we assume a homogeneous toroidal field at the base of the 
convection zone. Alternatively the magnetic field could be highly 
intermittent leading
to the following two complications: The Lorentz force is dependent on the
structure of the field in detail and cannot be expressed just by the mean 
field alone; the advection velocity of the mean field is not given by the mean 
flow, since field free plasma can flow around magnetic elements. In that sense 
for the case of intermittent field there are two extreme scenarios possible:
1. The field is highly intermittent and couples only weakly through the drag
force of individual flux elements to the mean flow. In this case the 
equatorward transport of field by the meridional flow is mainly dependent
on the strength of the coupling between mean field and mean flow (similar
to the behavior of individual flux tubes as discussed by 
\citet{Rempel:2003}); the change of the meridional flow pattern by the
feedback of the Lorentz force is a secondary effect.
2. The coupling between mean flow and mean field
is strong. In this case equatorward transport of field by the meridional flow
is dependent on the change of the meridional flow pattern caused by the Lorentz
force.

Even though we assume a homogeneous field in this study, our results also
apply to intermittent field in the case of strong coupling discussed above.
Independent from the particular way the 
field couples to the fluid, momentum conservation requires that the bulk force
acting on the bulk flow is given by the mean Lorentz force. However, the 
relation between the mean Lorentz force and the mean field strength can be 
more complicated as we illustrate in the following very simple example.

Assume the toroidal field consists of magnetic flux elements with total 
filling factor $f$ and field strength $B_0$. The mean tension force 
$\langle F_t\rangle$ is proportional to $f B_0^2$, while the mean field 
strength is given by $\langle B\rangle=f B_0$, therefore 
$\langle F_t\rangle\sim \langle B\rangle^2/f$. This means that for a given 
tension force the assumption of an intermittent field requires a factor of 
$f^{1/2}$ smaller mean field compared to the homogeneous case; 
however the field strength of individual flux elements would be a factor 
$f^{-1/2}$ larger.

In this paper we evaluate 
under which conditions an equatorward meridional flow at the base of the
convection zone can transport magnetic field equatorward. To this end we
compute at a fixed latitude the following effective transport velocity for
a given magnetic field configuration:
\begin{equation}
  v_{\theta}^{\rm eff}=\frac{\int v_{\theta}B r {\rm d}r}{\int B r {\rm d}r}\;,
  \label{transport}
\end{equation}
where $v_{\theta}^{\rm eff}>0$ means a transport of magnetic flux toward the 
equator. In the case of intermittent field $v_{\theta}^{\rm eff}$ will be an 
upper estimate for the transport capability.
Explicitly calculating the coupling between mean flow and mean field is 
beyond the scope of this paper and most likely highly dependent on the
exact field configuration. 

\section{Numerical results}
\begin{deluxetable}{ccccccccc}
  \tablecaption{Parameters of reference models\label{tab1}}
  \tablehead{ \colhead{model} & \colhead{$\Lambda_0$} & 
    \colhead{$\lambda$ [$\degr$]} & 
    \colhead{$\kappa_t$, $\nu_t$ [$\mbox{m}^2\,\mbox{s}^{-1}$]} }
  \startdata
  1 & $0.8$  & $15$   & $5\cdot 10^8$ \\
  2 & $1.49$ & $7.5$  & $5\cdot 10^8$ \\
  3 & $0.43$ & $22.5$ & $5\cdot 10^8$ \\
  4 & $1.36$ & $7.5$  & $2.5\cdot 10^8$ \\
  5 & $0.48$ & $22.5$ & $10^9$ \\
  \enddata
  \tablecomments{Parameters of differential rotation models used. Models
  1 to 3 show different meridional flow velocities for the same value of
  the turbulent diffusivity, whereas models 1, 4, and 5 show roughly the
  same meridional flow velocity for different values of the turbulent
  diffusivity. The parameter $\Lambda_0$ determines the amplitude of the
  turbulent angular momentum flux, the parameter $\lambda$ the direction
  of the angular momentum flux with respect to the axis of rotation.
  The quantities $\kappa_t$ and $\nu_t$ denote turbulent thermal diffusivity
  an turbulent viscosity, respectively.}
\end{deluxetable}
\begin{figure}
  \resizebox{\hsize}{!}{\includegraphics{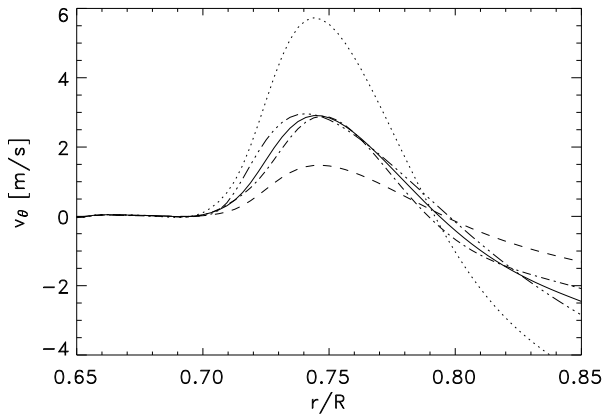}}
  \caption{Meridional flow velocity of the reference models listed in Table
    \ref{tab1}. Shown is a radial cut of the $\theta$ velocity at $30\degr$
    latitude of model 1 (solid), model 2 (dotted), model 3 (dashed), model 4
    (dashed-dotted), and model 5 (dashed-triple-dotted). Models 1 - 3 have 
    the same turbulent viscosity  but different flow velocities, while models
    4 and 5 have similar flow velocities as model 1, but different
    turbulent viscosities.
  }
  \label{f1}
\end{figure}
\begin{figure*}
  \resizebox{\hsize}{!}{\includegraphics{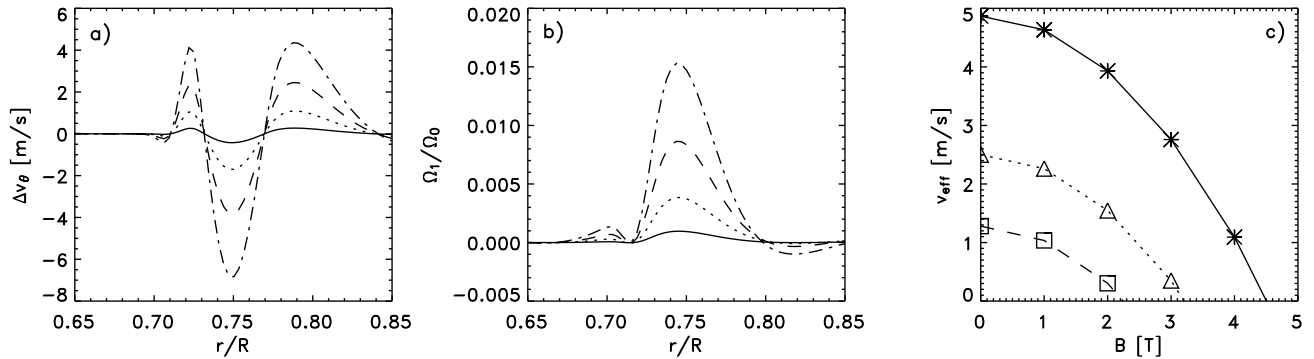}}
  \caption{Panel a): Changes of the meridional flow resulting from the 
    presence 
    of a toroidal field with an amplitude of $1\,\mbox{T}$ (solid), 
    $2\,\mbox{T}$ (dotted), $3\,\mbox{T}$ (dashed), and $4\,\mbox{T}$ 
    (dashed-dotted). The solutions shown here use the reference model 1;
    however, models 2 and 3 show a very similar flow pattern (the differences
    are of the order of a few percent). Panel b) shows the change of 
    differential rotation. The presence of a magnetic field leads to the 
    formation of a prograde flow within the
    magnetized region. The amplitude of this jet is strong enough to compensate
    around $70\%$ of the magnetic tension force through the resulting Coriolis
    force. Panel c) shows the effective transport velocity computed from
    Eq. (\ref{transport}) as function of field strength. Shown are the results
    for the reference models 1 (triangles), 2 (asterisks), and 3 (squares).
  }
  \label{f2}
\end{figure*}
\begin{figure*}
  \resizebox{\hsize}{!}{\includegraphics{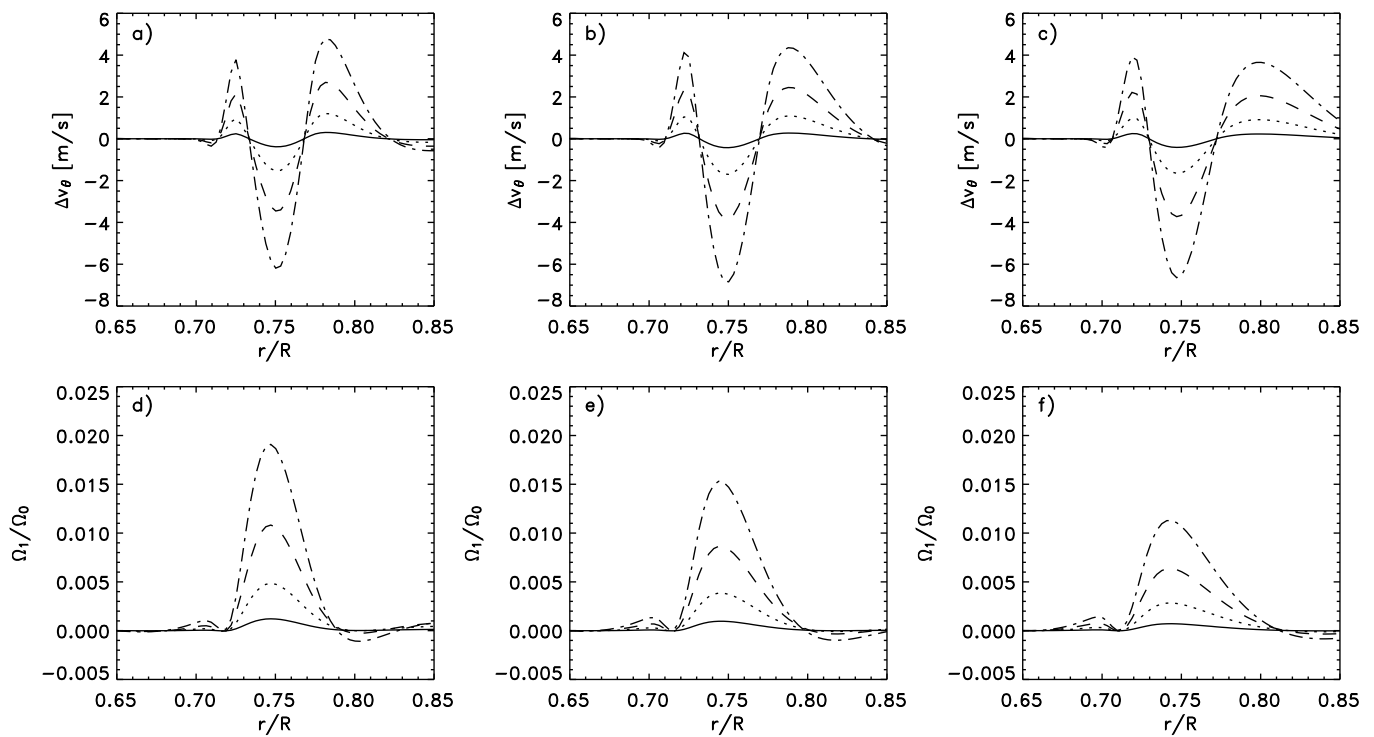}}
  \caption{Changes of meridional flow (top) and differential rotation (bottom)
    using the reference models 4, 1, and 5 (left to right). An increase of
    turbulent viscosity (left to right) while keeping the flow speed constant 
    leads to a spread of the side lopes and a reduction of the prograde flow 
    within the magnetized region. The linestyle indicates different magnetic 
    field strength as in Fig. \ref{f2}. The effective transport velocity is 
    only marginally affected (see Fig. \ref{f5}). 
  }
  \label{f3}
\end{figure*}
\begin{figure*}
  \resizebox{\hsize}{!}{\includegraphics{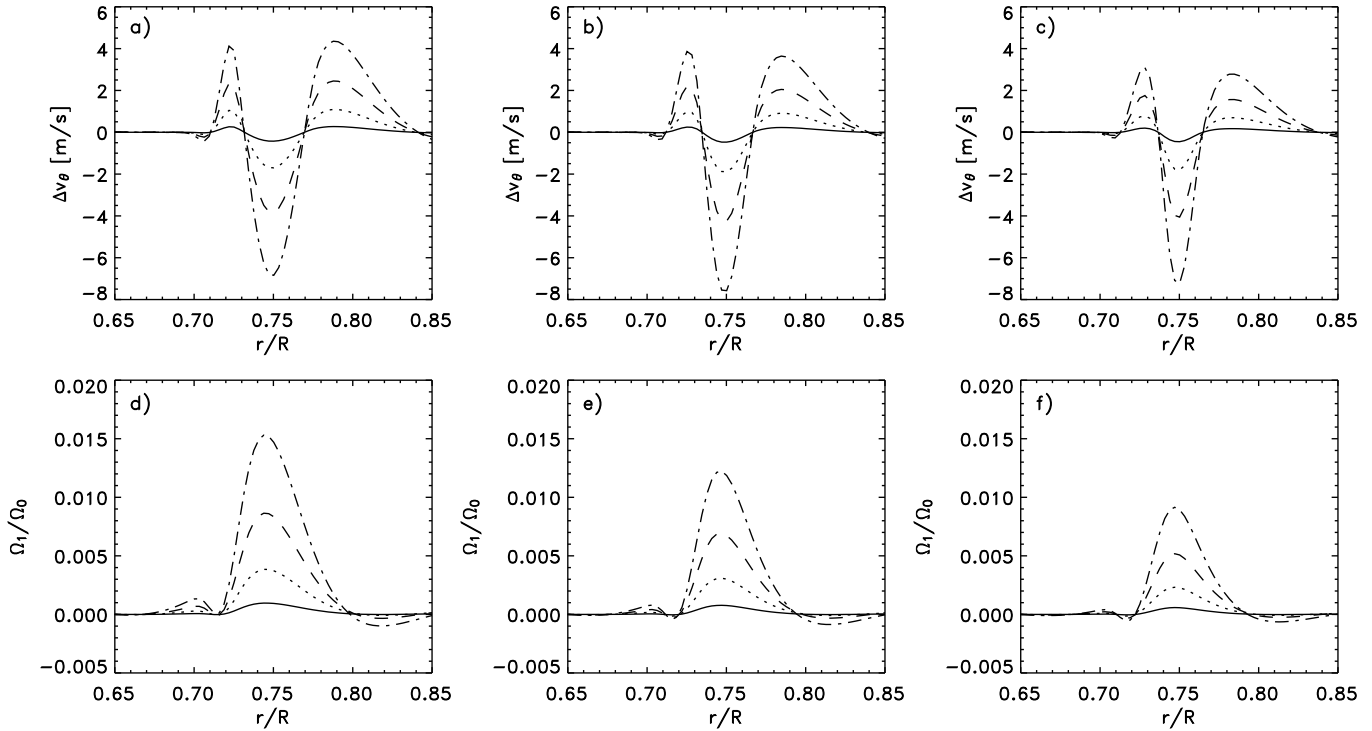}}
  \caption{Changes of meridional flow (top) and differential rotation (bottom)
    using the reference models 1 and a decreasing width of the magnetic field
    profile. The width of the field profile centered at $r=0.75 R_{\odot}$ 
    decreases from $0.05 R_{\odot}$ to $0.025 R_{\odot}$ (left to right). 
    Similar to the increase of turbulent viscosity, a
    decrease of width leads to a reduction of the prograde flow 
    within the magnetized region. The linestyle
    indicates different magnetic field strength as in Fig. \ref{f2}.
    The effective transport velocity is only 
    marginally affected (see Fig. \ref{f5}).
  }
  \label{f4}
\end{figure*}
\begin{figure}
  \resizebox{\hsize}{!}{\includegraphics{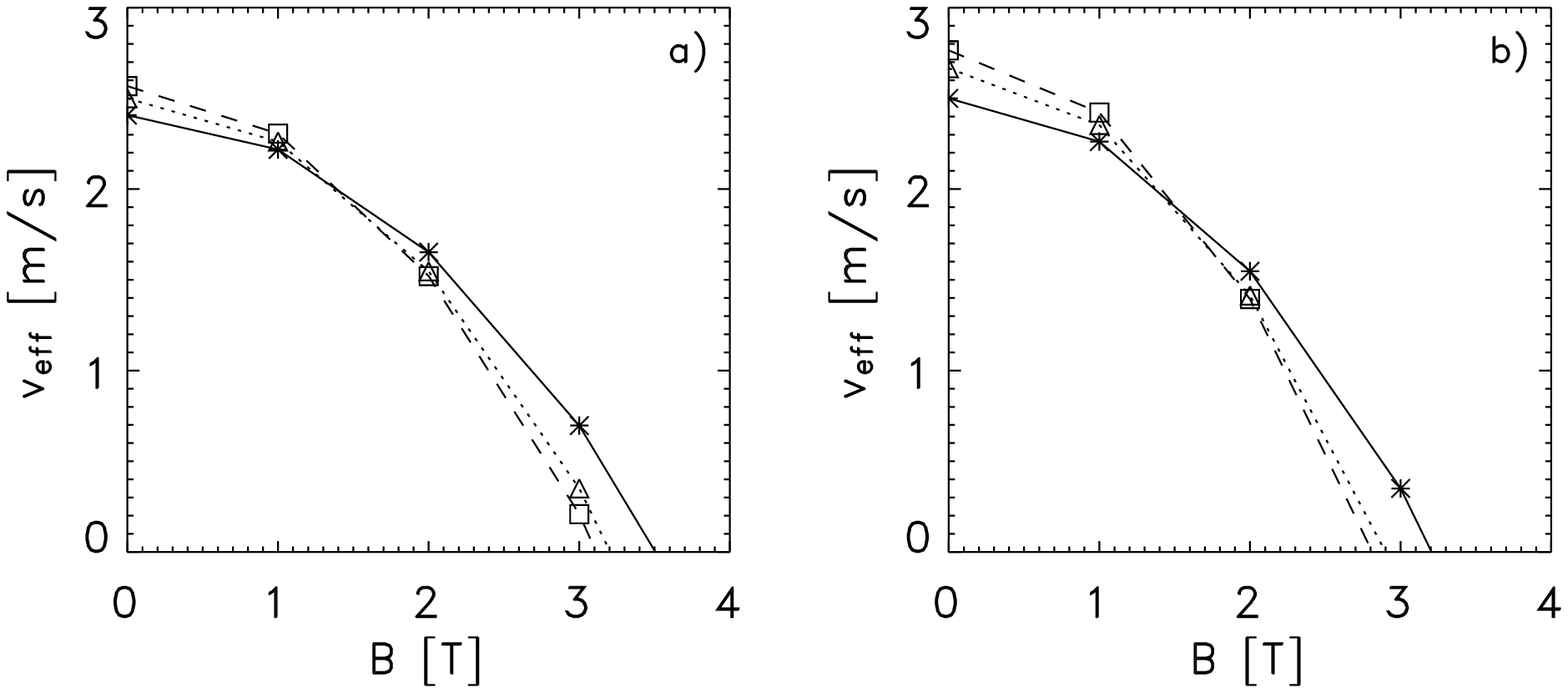}}
  \caption{Effective transport velocity for the cases shown in Fig. \ref{f3}
    (left) and Fig. \ref{f4} (right). Asterisks, triangles, squares 
    correspond to the solutions shown in Fig. \ref{f3} and \ref{f4} on the 
    left, middle, right, respectively.
  }
  \label{f5}
\end{figure}
In this section we use models for the differential rotation and 
meridional flow which are very similar to model 1 in \citet{Rempel:2005}.
Parameters that differ from those used in \citet{Rempel:2005} are 
summarized in Tab. \ref{tab1}. We have chosen the parameters such that 
we have three
models with meridional flow velocities varying by a factor of $4$, while the
turbulent diffusivity is unchanged (model 1, 2, 3) and three models with
a turbulent diffusivity varying by a factor of $4$, while the meridional
flow speed is roughly the same (model 1, 4, 5). Fig. \ref{f1} shows the flow
profile of $v_{\theta}$ at $30\degr$ latitude for the models listed in 
Tab. \ref{tab1}.

Fig. \ref{f2} summarizes the results obtained with the reference models
1, 2, and 3 (varying flow speed, but same turbulent viscosity). It turns
out that the changes of the meridional flow and differential rotation
are largely independent of the reference model (within a few percent 
variation).
Therefore we plot in Fig. \ref{f2} a) and b) the change of meridional
flow and differential rotation for the reference model 1. In each
case we add magnetic field with a maximum strength of $1\,\mbox{T}$, 
$2\,\mbox{T}$, $3\,\mbox{T}$, and $4\,\mbox{T}$. The magnetic 
field is located at $30\degr$ latitude and has a radial width of 
$0.05\,R_{\odot}$ and a parabolic profile in $B^2$ that is centered at
$0.75\,R_{\odot}$. In latitude we use a Gaussian profile with a width of 
about $15\degr$. For the case with $4\,\mbox{T}$ field strength the change
of meridional flow speed has a maximum amplitude of around 
$7\,\mbox{m}\,\mbox{s}^{-1}$
and is therefore larger than the meridional flow to be expected at the base
of the convection zone (the observed surface flow speed is around 
$10 - 20\,\mbox{m}\,\mbox{s}^{-1}$, which requires only a return flow of 
$1 - 2\,\mbox{m}\,\mbox{s}^{-1}$ at the base of the convection zone due to 
mass conservation). 

A consequence of the meridional flow change induced by the magnetic tension 
force is the formation of a prograde jet within the magnetized
region shown in Fig. \ref{f2} b), which partially compensates the magnetic
tension due to the Coriolis force. The amplitude of the jet that would be 
required for a complete compensation is given by (see section \ref{analytic} 
for a derivation):
\begin{equation}
  \frac{\Omega_1^{\prime}}{\Omega_0}=\frac{1}{2}\left(\frac{v_a}{\Omega_0 r 
    \sin\theta}\right)^2\;,
\end{equation}
with the Alfv\'en speed $v_a=B/\sqrt{\mu_0\rho}$.
For a field of $4\,\mbox{T}$ at $r=0.75\,R_{\odot}$, $30\degr$ latitude,
and value of $\rh_0=150\,\mbox{kg}\,\mbox{m}^{-3}$ this yields
$\Omega_1^{\prime}/\Omega_0\approx 0.025$, which means that for the case shown
in Fig \ref{f2} the jet compensates about $70\%$ of the magnetic tension force
(for the other field strengths shown, this ratio is about the same).

The interesting result is that the prograde jet forms independent of whether
the magnetic field is transported equatorward or poleward (angular momentum
conservation would only provide a prograde jet for a poleward movement). 
This is caused by the fact that the reference state (without any influence of
magnetic field) is characterized by an equilibrium between turbulent angular
momentum transport (including dissipative terms and non dissipative terms 
"$\Lambda$-effect") and the angular momentum transport of the meridional
flow (the divergence of the total angular momentum flux has to be zero for
a stationary solution). The magnetic tension force reduces the equatorward
flow speed at the base of the convection zone, disturbing this balance and 
therefore forcing a change of the differential rotation. Formally
this result can be understood as follows: Let 
$\vec{F}_{\Lambda}^0$, $\vec{F}_{\nu}^0$, and $\vec{F}_{m}^0$ 
be the angular momentum fluxes in the 
reference state due to turbulent non-dissipative angular momentum transport 
($\Lambda$-effect), turbulent dissipation, and meridional flow, respectively. 
Then the stationary reference state is characterized by 
$\vec{\nabla}\cdot(\vec{F}_{\Lambda}^0+\vec{F}_{\nu}^0+\vec{F}_{m}^0)=0$. 
The presence 
of a magnetic field changes the the meridional flow, leading to a perturbation 
$\vec{F}_{m}^1$, which changes the differential rotation and leads to a
perturbation of the dissipative angular momentum flux $\vec{F}_{\nu}^1$. 
Since the $\Lambda$-effect remains unaffected in this case, the new stationary 
equilibrium requires $\vec{\nabla}\cdot(\vec{F}_{m}^1+\vec{F}_{\nu}^1)=0$. 
Since for all cases considered $\vec{F}_{m}^1$ is always poleward directed 
in the magnetized region, this requires the formation of a prograde jet, even 
though $\vec{F}_{m}^0+\vec{F}_{m}^1$ can be still equatorward directed.

\citet{Moreno-Insertis:etal:1992,Rempel:etal:2000} discussed the formation
of prograde jets required for equilibrium states of toroidal field at the base 
of the convection zone. In their studies a poleward movement of the magnetic 
field was always required, since turbulent angular momentum transport and
meridional flow were not considered in the reference state. The magnetic
field strength of around $10\,\mbox{T}$ they considered is also 
significantly larger than the field strength we will focus on in this paper.

While the changes of the meridional flow are independent
of the meridional flow speed in the reference model, the question whether
magnetic field is still transported equatorward depends on the total flow
speed. Fig. \ref{f2} c) shows the effective transport velocity according to 
Eq. (\ref{transport}) for the three cases. In the case of a meridional flow
of around $1.5\,\mbox{m}\,\mbox{s}^{-1}$, magnetic fields up to a 
about $2.2\,\mbox{T}$ can be transported equatorward, while a 
$6\,\mbox{m}\,\mbox{s}^{-1}$ flow could transport up to $4.5\,\mbox{T}$ 
equatorward.
Note that the critical field strength scales roughly with the square-root of
the meridional flow velocity. This is different from the case of individual
flux tubes coupling to the flow through the drag force, which results in
$B\sim v$.

Fig. \ref{f3} shows the result for reference models 1, 4, and 5
that have roughly the same flow velocity, but a variation in the turbulent 
viscosity by a factor of 4. Even though the amplitude of the prograde jet 
changes significantly, the transport capability of the meridional flow
(see Fig. \ref{f5}) is only marginally affected.
In the model 4 with a value of $\nu_t=2.5\cdot 10^8\mbox{m}^2\,\mbox{s}^{-1}$
the jet compensates around $90\%$ of the curvature force (left panels),
whereas in  model 5 with a value of $\nu_t=10^9\mbox{m}^2\,\mbox{s}^{-1}$
the jet compensates only around $50\%$ of the curvature force (right panels).
This does not affect strongly the transport capability of the meridional flow
since at the same time with increasing turbulent viscosity the transport of 
magnetic
field through viscous drag increases (a typical value of the Reynolds number
on scales of the magnetic field is of the order of one due to the small
meridional flow velocity and the large turbulent viscosity). We will discuss
this in detail in section \ref{analytic}.

Fig. \ref{f4} shows the result for different radial widths of the magnetic 
field. In all cases we use model 1 as the reference model. Panels a) and d)
show results for a width of $0.05\,R_{\odot}$, panels b) and e) for a width of 
$0.035\,R_{\odot}$, and panels c) and f) for a width of $0.025\,R_{\odot}$.
A comparison with Fig. \ref{f3} shows clearly that a reduction of the width
of the magnetic field is equivalent to an increase of the viscosity as expected
from a simple scaling argument.

Fig. \ref{f5} shows the effective transport velocity for the cases shown in
Fig. \ref{f3} (left panel) and Fig. \ref{f4} (right panel). Changes of the
turbulent viscosity (by a factor of 4) and the width of the magnetic field
(by a factor of 2) considered here influence the critical field strength
up to which an equatorward transport of flux is possible by around $10\%$.

\section{Analytic estimate}
\label{analytic}
Let $v_r$, $v_{\theta}$, and $\Omega_1$ be a stationary solution of the
differential rotation problem without a magnetic field, and $v_r^{\prime}$,
$v_{\theta}^{\prime}$, and $\Omega_1^{\prime}$ perturbations around that state
caused by the presence of the magnetic field. Only considering the 
$\theta$-component of the momentum equation, we can estimate the balance 
between Coriolis force, magnetic tension and viscous stress by:
\begin{equation}
  2\Omega_0\Omega_1^{\prime}r\sin\theta\cos\theta
  -\frac{B^2}{\mu_0\rh_0r}\cot\theta-c_d\,\nu_t\frac{v_{\theta}^{\prime}}{d^2}
  =0\;,
  \label{est1}
\end{equation}
where $d$ denotes a length associated with the radial width of the magnetic
field and $c_d$ is a coefficient, which takes care of the more complicated
flow structure. The formulation given here relates 
to the more general formulation of the drag force $\sim c_w\rh v^2/(2\,d)$ 
through the assumption $c_w\sim c_d/Re$ with the turbulent Reynolds number
$Re=v\,d/\nu_t$. Since the large scale flow pattern in combination with the 
large turbulent viscosity is a flow with a small turbulent Reynolds number 
$\sim 1$, it can be expected that the meridional flow behaves like a highly 
viscous fluid. This is justified as long as the scale of the flow field is 
larger than the typical turbulence scale. In this limit typical values for
$c_d$ should be of the order of $10$ (see text books on fluid dynamics for 
Stokes law). For small magnetic flux tubes a 
formulation with $c_w={\rm const}$ rather than $c_d={\rm const}$ is more valid.
We show later that for a reasonable choice of $c_d$ a good agreement between 
this analytic scaling analysis and the numerical result can be achieved.

We emphasize that drawing parallels between laminar, viscid 
laboratory flows and highly turbulent astrophysical flows with a low turbulent 
Reynolds number is speculative; however it is unavoidable
when applying the mean field approach, leading to the parameterization of large
turbulent viscosities.

The balance between angular momentum transport and viscous dissipation yields: 
\begin{equation}
  -2\Omega_0\frac{v_{\theta}^{\prime}}{r}\cot\theta
  -\nu_t\frac{\Omega_1^{\prime}}{d^2}=0\label{est2}
\end{equation}
We did not introduce here an additional free parameter in front of the
diffusive loss term, since this prefactor $\sim1$ can be easily absorbed into
the definition of the length scale $d$. Combining both equations gives a 
relation between $B$ and $v_{\theta}^{\prime}$:
\begin{equation}
  B^2=-\mu_0\rh_0 r \tan\theta\left((2\Omega_0\cos\theta)^2\frac{d^2}{\nu_t}
  +c_d\,\frac{\nu_t}{d^2}\right)v_{\theta}^{\prime}\label{est3}
\end{equation}
An equatorward transport of the magnetic field requires 
$\vert v_{\theta}^{\prime}\vert \lesssim \vert v_{\theta}\vert$, which yields 
an upper limit for magnetic field strength of:
\begin{equation}
  B^2\sim\mu_0\rh_0 r \tan\theta\left((2\Omega_0\cos\theta)^2
  \frac{d^2}{\nu_t}+c_d\,\frac{\nu_t}{d^2}\right)v_{\theta}\label{est4}
\end{equation}
The importance of the two terms in the angular brackets depends on the value
of the turbulent viscosity $\nu_t$ and the width of the magnetic field $d$.
For a given width, the first term is important for
low viscosity, the second one for high viscosity. Both are of the same 
importance if
\begin{equation}
  \nu_t=\nu_{\rm crit}=\frac{2}{c_d^{1/2}}\Omega_0 d^2\cos\theta\;.
  \label{est5}
\end{equation}
Eq. (\ref{est4}) can be written together with Eq. (\ref{est5}) in the form:
\begin{equation}
  B\sim 2^{1/2} c_d^{1/4}\sqrt{\mu_0\rh_0\Omega_0 r \sin\theta v_{\theta}}
  \sqrt{\frac{\nu_{\rm crit}}{\nu_t}+\frac{\nu_t}{\nu_{\rm crit}}}
  \label{est6}
\end{equation}
Eq. (\ref{est5}) shows clearly the relation $B\sim v_{\theta}^{1/2}$ as 
suggested by Fig. \ref{f2}c.

Inserting Eq. (\ref{est3}) into Eq. (\ref{est2}) yields for the perturbation
of $\Omega$:
\begin{equation}
  \frac{\Omega_1^{\prime}}{\Omega_0}=\frac{1}{2}\left(\frac{v_a}{\Omega_0 r 
    \sin\theta}\right)^2 \left[1+\left(\frac{\nu_t}{\nu_{\rm crit}}\right)^2
    \right]^{-1}\label{est7}
\end{equation}

In the case of very low viscosity the perturbation of $\Omega$ is given by 
\begin{equation}
  \frac{\Omega_1^{\prime}}{\Omega_0}=\frac{1}{2}\left(\frac{v_a}{\Omega_0 r 
    \sin\theta}\right)^2\;,\label{est8}
\end{equation}
which is exactly the value required to balance the magnetic tension through
the Coriolis force. In the case of large viscosity the Coriolis force is
unimportant and the magnetic field is dragged by the meridional flow through
viscous coupling.

In order to compare these estimates to the numerical results we use the model
1, which is shown in Fig. \ref{f2} panels a) and b). As already mentioned
earlier the jet amplitude is sufficient to compensate for $70\%$ of the 
curvature stress, which yields according to Eq. (\ref{est7}) 
$\nu_t\sim 0.65 \nu_{\rm crit}$. With a value of 
$\nu_t\sim 2.5\cdot 10^8\mbox{m}^2\,\mbox{s}^{-1}$ at $0.75\,R_{\odot}$
(roughly half the convection zone value) and 
$\Omega_0=2.7\cdot 10^{-6}\,\mbox{s}^{-1}$ Eq. (\ref{est5}) yields
$d=c_d^{1/4} 0.014\,R_{\odot}$. Using a value of $c_d=10$ yields
$d=0.025\,R_{\odot}$, which is half of assumed width of the magnetic field.
With $c_d=10$, $\nu_t\sim 0.65 \nu_{\rm crit}$, 
$\rh_0=150\,\mbox{kg}\,\mbox{m}^{-3}$, $r=0.75\,R_{\odot}$, and
$v_{\theta}\sim 2.5\,\mbox{m}\,\mbox{s}^{-1}$ 
Eq. (\ref{est6}) yields $B\sim 2.8\,\mbox{T}$, which is very close to the 
numerical result. 

Since Eq. (\ref{est6}) as function of $\nu_t/\nu_{\rm crit}$ has a local
minimum for $\nu_t=\nu_{\rm crit}$, the dependence of $B$ on $\nu_t$ and
$d$ (through $\nu_{\rm crit}$, Eq. (\ref{est5})) is expected to be rather weak,
whereas $\Omega_1^{\prime}/\Omega_0$ shows a much stronger dependence as found
in Figs. \ref{f3} and \ref{f4}. In more detail, Eq. (\ref{est7})
does not exactly reflect the scaling of $\Omega_1^{\prime}/\Omega_0$ with
$\nu_t$ and $d$ found in Figs. \ref{f3} and \ref{f4}, which suggests that
$c_d$ is depending on $\nu_t$ and $d$ itself, however the tendency is
indicated correctly. 

\section{Solutions with quenched viscosity}
\begin{figure*}
  \resizebox{\hsize}{!}{\includegraphics{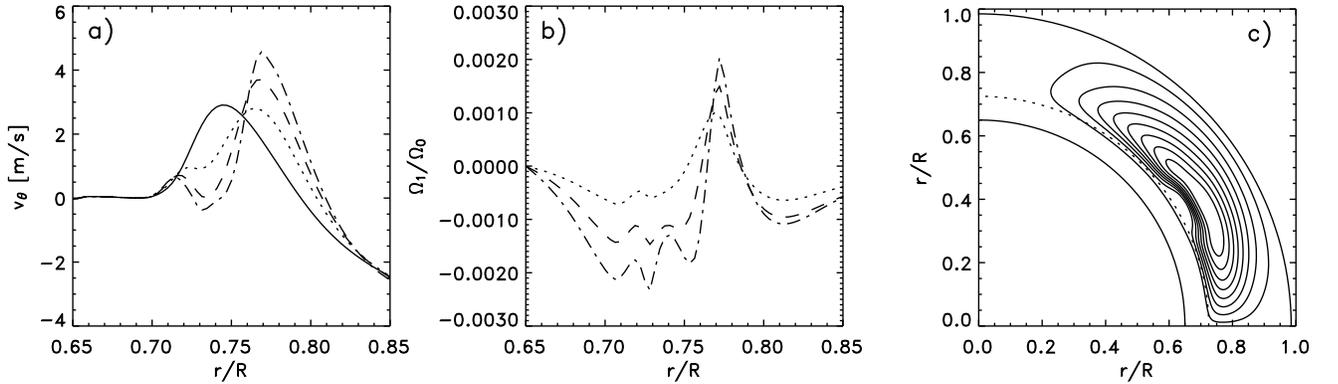}}
  \caption{Solution with a quenched viscosity. Panel a) shows the
    the meridional flow speed, where the dotted line indicates a
    quenching by a factor of $2$, the dashed line a quenching by a factor of 
    $4$, and the dashed-dotted line a quenching by a factor of $8$. The solid 
    line indicates the reference solution without any influence of the 
    magnetic field. Panel b) shows the change of differential rotation. Unlike
    the cases shown in Figs. \ref{f2} - \ref{f4} there is no formation of a 
    significant jet. Panel c) shows the streamlines for the case with a 
    quenching of the viscosity by a factor of $8$.
  }
  \label{f6}
\end{figure*}
So far we have discussed the back-reaction of toroidal field on the meridional 
low through the magnetic curvature force. Since the equipartition field 
strength at the base of the convection zone is around $1\,\mbox{T}$ (based on 
mixing-length models) the magnetic field also quenches the turbulent viscosity 
to some extent. Since this also changes the turbulent angular momentum flux 
an influence on the meridional flow is expected. In order to demonstrate
this effect use a very weak field of $1\,\mbox{T}$ field strength, which has   
nearly no influence on the meridional flow through the tension force, and 
include a quenching of the viscosity given by
\begin{eqnarray}
  \nu_t^{\prime}=\nu_t\left[1+\left(\frac{B}{B_{\rm eq}}\right)^2\right]^{-1}
  \;.
\end{eqnarray}
Note that $\nu_t$ scales in our model the viscous stress and the turbulent 
angular momentum transport ($\Lambda$-effect).
Fig. \ref{f5} a) shows the results for a the values 
$B_{\rm eq}=B_{\rm max}$, $B_{\rm eq}=B_{\rm max}/\sqrt{3}$, and 
$B_{\rm eq}=B_{\rm max}/\sqrt{7}$
that correspond to a quenching by a factor of $2$, $4$, and $8$, respectively.
A comparison with the results presented in Figs. \ref{f2} to \ref{f4} shows
that the back-reaction through quenching of viscosity has, at least for
weak field, a much stronger effect than the direct feedback through magnetic
tension. Whereas in the case of feedback through magnetic tension the 
meridional flow moves around the magnetized region on both sides (below and
above), in the case of quenching the meridional flow closes above the 
magnetized region. The changes of the differential rotation (panel b) are
around one order of magnitude lower than in Figs. \ref{f2} to \ref{f4} and
no distinct jet forms. 

This results from the fact that the change in the viscosity
changes the parametrized turbulent angular momentum transport that is the
indirect driver for the meridional flow. In our model the meridional return
flow is localized in the region where the turbulent viscosity shows the
strongest radial gradient. The quenching of the viscosity moves therefore
the return flow upward, as clearly indicated in Fig. \ref{f6} panel c). 
However, the transport of magnetic flux is not switched off
completely in this case. The magnetic field we assumed is centered at
$r=0.75\,R_{\odot}$ and has a width of $0.05\,R_{\odot}$, which means that
the upper half is still in a region of considerable flow speed. For example
the effective transport velocity is only reduced by a factor of two when 
the turbulent viscosity is quenched by a factor of $8$).
 
\section{Implications for solar dynamo models}
This investigation shows as a very robust result that the transport capability 
of the meridional flow is mainly determined by the flow velocity. For a flow
velocity of the return flow at the base of the convection zone of about 
$2.5\,\mbox{m}\,\mbox{s}^{-1}$ the maximum field strength that can be 
transported is around $3\,\mbox{T}$ ($30\,\mbox{kG}$). 
This field strength is found to be rather insensitive to the turbulent 
viscosity or width of the magnetic band. Inspecting
Eq. (\ref{est6}) shows that a larger value for this field strength would
require either a very small or a very large value of $\nu_t/\nu_{\rm crit}$
or a very large value of $c_d$.
Since the meridional return flow is located roughly where $\nu_t$ shows the
strongest radial gradient at the base of the convection zone the value of
$\nu_t$ in that region will reflect more convection zone values rather than
very small overshoot values. The value of $\nu_{\rm crit}$ is mainly influenced
through the effective thickness $d$. Values much larger than the value used
in this investigation are not feasible, since then the magnetic layer would
have a larger extent in radius than the meridional return flow. For very small
values of $d$ an increase in the magnetic field strength that can be 
transported is
expected. This is not too surprising, since for magnetic flux tubes the 
influence of the drag force is anti proportional to the diameter. 
\citet{Rempel:2003} showed that flux tubes with a diameter of less than 
$100\,\mbox{km}$ can be transported equatorward even if the field strength 
is around $10\,\mbox{T}$, however the magnetic flux associated with these 
flux tubes is orders of magnitude smaller than the flux of a typical sunspot.
We want to emphasize that for flux tubes the use of $c_d\sim Re$
($c_w=\mbox{const}$) gives a different scaling of $B\sim v/d^{1/2}$ 
compared to the $B\sim (\nu_t v)^{1/2}/d$ scaling derived
from Eq. (\ref{est6}) in the limit of small values for $d$.

Feedback through quenching of turbulent viscosity (also affecting the 
$\Lambda$-effect, which is $\sim \nu_t$ in our model) leads to a significant
modification of the meridional flow if the field strength exceeds 
equipartition. Since typical mixing-length estimates for $B_eq$ at the base 
of the convection zone are around $1\,\mbox{T}$, this happens in the same 
field strength range where the direct feedback through magnetic tension becomes
important, too. Therefore if we consider both effects together the effective
transport velocities are modified, but the results do not change dramatically.

To summarize, the result that the consideration of the magnetic curvature force
limits the magnetic field strength that can be transported towards the equator
to about $3\,\mbox{T}$ ($30\,\mbox{kG}$) seems to be very robust within the
framework of the mean field model used in this investigation.

Current kinematic solar flux-transport dynamos models rely 
on the equatorward transport of toroidal field at the base of the convection
through the meridional flow (see comparison of solutions with and without
meridional flow in \citet{Dikpati:Charbonneau:1999,Dikpati:Gilman:2001b}).
The meridional flow velocity at the base of the convection zone assumed in
most of these models is around $1 - 2\,\mbox{m}\,\mbox{s}^{-1}$, which is close
to the values considered in this paper. That value is also consistent with 
the observed surface flow of around $10 - 20\,\mbox{m}\,\mbox{s}^{-1}$ and a 
dynamo period around $22$ years.
The main consequence of the work presented here for flux-transport dynamos 
can be summarized as follows:
Any field exceeding a few $\mbox{T}$ ($10\,\mbox{kG}$) cannot be transported
toward the equator through a meridional flow with an amplitude of a few 
$\mbox{m}\,\mbox{s}^{-1}$. If the solar dynamo produces stronger field
(e.g. $100\,\mbox{kG}$ as inferred from studies of rising magnetic flux tubes
\citep{Choudhuri:Gilman:1987,Fan:etal:1993,
Schuessler:etal:1994,Caligari:etal:1995,Caligari:etal:1998}), 
the field must get amplified locally through induction effects. The stronger 
field could be in an equilibrium as discussed by 
\citet{Moreno-Insertis:etal:1992,Rempel:etal:2000,Rempel:Dikpati:2003}, which
would prevent further poleward movement, but an equatorward transport is not
possible.

First attempts to include the feedback on meridional flow and differential
rotation in a 'dynamic' dynamo model \citep{Rempel:etal:2005} show that
flux-transport dynamos work with toroidal field strengths up to around
$30\,\mbox{kG}$, however additional constraints apply through the observed
limits on the amplitude of torsional oscillations 
\citep{Howe:etal:2004,Rempel:etal:2005}, if the feedback on differential 
rotation is included, too.

The problem addressed in this paper is not limited to flux-transport dynamos.
Any dynamo will face the problem that the toroidal field will start moving
toward the pole due to the magnetic tension force if the field strength is
large enough. If the propagation of the magnetic activity is not an advection
effect, but a classic dynamo wave, the wave will have to compete with the 
poleward movement induced by the magnetic tension force. Since the changes
of the meridional flow shown in Fig. \ref{f2} are only weakly dependent on the
meridional flow speed of the reference state, they also give an estimate how 
large the poleward movement would be if no meridional flow is present.

We have shown that the formation of a prograde jet is an unavoidable 
consequence when including the magnetic tension force. Recently 
\citet{Dalsgaard:etal:2004} tried to detect jets associated with toroidal
magnetic field in the tachocline. The detection limit they found is of the
order of $2-4$ nHz, which is around $1\%$ of the rotation rate. The jets we
see in this study have an amplitude of around  $1.5\%$ for the strongest field
we considered ($4\mbox{T}$). Magnetic field of $2\mbox{T}$ or less leads to the
formation of jets with less than $0.5\%$ amplitude. These jets are therefore at
or below the detection limit of current helioseismic techniques. If the 
magnetic field has a more complicated intermittent and also non-axisymmetric 
structure it is likely that the amplitude of the prograde jet is lower than
predicted by our axisymmetric model. 

\acknowledgements
Stimulating discussions and helpful comments on a draft version of this paper
by Mausumi Dikpati, Keith MacGregor and Peter Gilman are gratefully 
acknowledged. The author thanks the anonymous referee for a very helpful 
review. 

\bibliographystyle{natbib/apj}
\bibliography{natbib/apj-jour,natbib/papref}

\begin{thebibliography}{27}
\expandafter\ifx\csname natexlab\endcsname\relax\def\natexlab#1{#1}\fi

\bibitem[{{Braun} \& {Fan}(1998)}]{Braun:Fan:1998}
{Braun}, D.~C. \& {Fan}, Y. 1998, \apjl, 508, L105

\bibitem[{{Brun} \& {Toomre}(2002)}]{Brun:Toomre:2002}
{Brun}, A.~S. \& {Toomre}, J. 2002, \apj, 570, 865

\bibitem[{{Caligari} {et~al.}(1995){Caligari}, {Moreno-Insertis}, \&
  {Sch\"ussler}}]{Caligari:etal:1995}
{Caligari}, P., {Moreno-Insertis}, F., \& {Sch\"ussler}, M. 1995, \apj, 441,
  886

\bibitem[{{Caligari} {et~al.}(1998){Caligari}, {Sch\"ussler}, \&
  {Moreno-Insertis}}]{Caligari:etal:1998}
{Caligari}, P., {Sch\"ussler}, M., \& {Moreno-Insertis}, F. 1998, \apj, 502,
  481

\bibitem[{{Choudhuri} \& {Gilman}(1987)}]{Choudhuri:Gilman:1987}
{Choudhuri}, A.~R. \& {Gilman}, P.~A. 1987, \apj, 316, 788

\bibitem[{{Choudhuri} {et~al.}(1995){Choudhuri}, {Sch\"ussler}, \&
  {Dikpati}}]{Choudhuri:etal:1995}
{Choudhuri}, A.~R., {Sch\"ussler}, M., \& {Dikpati}, M. 1995, \aap, 303, L29

\bibitem[{{Christensen-Dalsgaard} {et~al.}(2004){Christensen-Dalsgaard},
  {Corbard}, {Dikpati}, {Gilman}, \& {Thompson}}]{Dalsgaard:etal:2004}
{Christensen-Dalsgaard}, J., {Corbard}, T., {Dikpati}, M., {Gilman}, P.~A., \&
  {Thompson}, M.~J. 2004, in ESA SP-559: SOHO 14 Helio- and Asteroseismology:
  Towards a Golden Future, 376--+

\bibitem[{{Dikpati} \& {Charbonneau}(1999)}]{Dikpati:Charbonneau:1999}
{Dikpati}, M. \& {Charbonneau}, P. 1999, \apj, 518, 508

\bibitem[{{Dikpati} \& {Gilman}(2001)}]{Dikpati:Gilman:2001b}
{Dikpati}, M. \& {Gilman}, P.~A. 2001, \apj, 559, 428

\bibitem[{{Fan} {et~al.}(1993){Fan}, {Fisher}, \& {DeLuca}}]{Fan:etal:1993}
{Fan}, Y., {Fisher}, G.~H., \& {DeLuca}, E.~E. 1993, \apj, 405, 390

\bibitem[{{Gilman} \& {Miller}(1986)}]{Gilman:Miller:1986}
{Gilman}, P.~A. \& {Miller}, J. 1986, \apjs, 61, 585

\bibitem[{{Glatzmaier} \& {Gilman}(1982)}]{Glatzmaier:Gilman:1982}
{Glatzmaier}, G.~A. \& {Gilman}, P.~A. 1982, \apj, 256, 316

\bibitem[{{Haber} {et~al.}(2002){Haber}, {Hindman}, {Toomre}, {Bogart},
  {Larsen}, \& {Hill}}]{Haber:etal:2002}
{Haber}, D.~A., {Hindman}, B.~W., {Toomre}, J., {Bogart}, R.~S., {Larsen},
  R.~M., \& {Hill}, F. 2002, \apj, 570, 855

\bibitem[{{Howe} {et~al.}(2004){Howe}, {Rempel}, {Christensen-Dalsgaard},
  {Hill}, {Komm}, {Schou}, \& {Thompson}}]{Howe:etal:2004}
{Howe}, R., {Rempel}, M., {Christensen-Dalsgaard}, J., {Hill}, F., {Komm},
  R.~W., {Schou}, J., \& {Thompson}, M.~J. 2004, in ESA SP-559: SOHO 14 Helio-
  and Asteroseismology: Towards a Golden Future, 468--+

\bibitem[{{K{\" u}ker} \& {Stix}(2001)}]{Kueker:Stix:2001}
{K{\" u}ker}, M. \& {Stix}, M. 2001, \aap, 366, 668

\bibitem[{{Kitchatinov} \& {R\"udiger}(1993)}]{Kitchatinov:Ruediger:1993}
{Kitchatinov}, L.~L. \& {R\"udiger}, G. 1993, \aap, 276, 96

\bibitem[{{Kitchatinov} \& {R\"udiger}(1995)}]{Kitchatinov:Ruediger:1995}
---. 1995, \aap, 299, 446

\bibitem[{{Miesch} {et~al.}(2000){Miesch}, {Elliott}, {Toomre}, {Clune},
  {Glatzmaier}, \& {Gilman}}]{Miesch:etal:2000}
{Miesch}, M.~S., {Elliott}, J.~R., {Toomre}, J., {Clune}, T.~L., {Glatzmaier},
  G.~A., \& {Gilman}, P.~A. 2000, \apj, 532, 593

\bibitem[{{Moreno-Insertis} {et~al.}(1992){Moreno-Insertis}, {Sch\"ussler}, \&
  {Ferriz-Mas}}]{Moreno-Insertis:etal:1992}
{Moreno-Insertis}, F., {Sch\"ussler}, M., \& {Ferriz-Mas}, A. 1992, \aap, 264,
  686

\bibitem[{{Rempel}(2003)}]{Rempel:2003}
{Rempel}, M. 2003, \aap, 397, 1097

\bibitem[{{Rempel}(2005)}]{Rempel:2005}
---. 2005, \apj, 622, 1320

\bibitem[{{Rempel} \& {Dikpati}(2003)}]{Rempel:Dikpati:2003}
{Rempel}, M. \& {Dikpati}, M. 2003, \apj, 584, 524

\bibitem[{{Rempel} {et~al.}(2005){Rempel}, {Dikpati}, \&
  {MacGregor}}]{Rempel:etal:2005}
{Rempel}, M., {Dikpati}, M., \& {MacGregor}, K. 2005, ESA SP-560, CS13
  Proceedings, 913

\bibitem[{{Rempel} {et~al.}(2000){Rempel}, {Sch{\" u}ssler}, \& {T{\'
  o}th}}]{Rempel:etal:2000}
{Rempel}, M., {Sch{\" u}ssler}, M., \& {T{\' o}th}, G. 2000, \aap, 363, 789

\bibitem[{{R\"udiger} {et~al.}(1998){R\"udiger}, {von Rekowski}, {Donahue}, \&
  {Baliunas}}]{Ruediger:etal:1998}
{R\"udiger}, G., {von Rekowski}, B., {Donahue}, R.~A., \& {Baliunas}, S.~L.
  1998, \apj, 494, 691

\bibitem[{{Sch\"ussler} {et~al.}(1994){Sch\"ussler}, {Caligari}, {Ferriz-Mas},
  \& {Moreno-Insertis}}]{Schuessler:etal:1994}
{Sch\"ussler}, M., {Caligari}, P., {Ferriz-Mas}, A., \& {Moreno-Insertis}, F.
  1994, \aap, 281, L69

\bibitem[{{Zhao} \& {Kosovichev}(2004)}]{Zhao:Kosovichev:2004}
{Zhao}, J. \& {Kosovichev}, A.~G. 2004, \apj, 603, 776

\end{thebibliography}

%\bibliography{ms}

\end{document}